\documentclass[ prl, twocolumn,   superscriptaddress]{revtex4-1} %twocolumn/preprint,  groupedaddress/superscriptaddress, longbibliography, linenumbers,
\usepackage{graphicx}
\usepackage{hyperref}
\usepackage{epstopdf}			% Use eps figures
\usepackage{color}
\usepackage{amsmath} % split environment
\usepackage{mathtools} % multlined environment
\usepackage{ctable}
\usepackage{tabularx}

\newcommand{\kT}{k_\mathrm{B}T}

\begin{document}
%\linenumbers
%\lipsum[1-5]
%\begin{widetext}
%======================================
% RevTeX
\title{Power-Law Stretching of Associating Polymers in Steady-State Extensional Flow }  
\date{\today}
\author{Charley Schaefer}
\email{charley.schaefer@york.ac.uk}
\affiliation{Department of Physics, University of York, Heslington, York, YO10 5DD, UK}
%\author{Peter R.  Laity}
%\email{petelaity@aol.com}
%\affiliation{Department of Materials Science and Engineering, The University of Sheffield, Sir Robert Hadfield Building, Mappin Street, Sheffield, S1 3JD, UK}
%\author{Chris Holland}
%\email{christopher.holland@sheffield.ac.uk}
%\affiliation{Department of Materials Science and Engineering, The University of Sheffield, Sir Robert Hadfield Building, Mappin Street, Sheffield, S1 3JD, UK}
\author{Tom C. B. McLeish}
%\email{tom.mcleish@york.ac.uk}
\affiliation{Department of Physics, University of York, Heslington, York, YO10 5DD, UK}
%======================================

% \pacs{}

% UK-US:
% bre-ber (fibre, centre - fiber, center)
% ise-ize
% or-our (color, neighbor - colour, neighbour

%\begin{tocentry}
{
%test tocentry
% \includegraphics*[height=3.0cm]{TOC_morph.png}
% \includegraphics*[trim=0cm 20.6cm 8cm 0.3cm, height=3.5cm  ]{TOC_scaling.eps}\\
%test tocentry
% \includegraphics*[trim=0cm 0.0cm 0cm 0.0cm, height=3.5cm  ]{fig/TOC.eps}
}
%\end{tocentry}

\begin{abstract}
{  
  We present a tube model for the Brownian dynamics of associating polymers in extensional flow. In linear response,
the model confirms the analytical predictions for the ‘sticky diffusivity’ by Leibler-Rubinstein-Colby theory. Although a
single-mode DEMG approximation accurately describes the transient stretching of the polymers above a `sticky'
Weissenberg number (product of the strain rate with the sticky-Rouse time), the pre-averaged model fails to capture
a remarkable development of a power-law distribution of stretch in steady-state extensional flow: while the mean
stretch is finite, the fluctuations in stretch may diverge. We present an analytical model that shows how strong
stochastic forcing drive the long tail of the distribution,  gives rise to ‘rare events’ of reaching a threshold stretch and constitutes a framework within which nucleation rates of flow-induced crystallization may understood in systems of associating polymers under flow. The model also exemplifies a wide class of driven
systems possessing strong, and scaling, fluctuations.
}
\end{abstract}

\maketitle
%\tableofcontents
%
%
\newpage

%\section{Introduction}

  The natural or artificial production of high-performance polymeric materials requires precise control over flow-induced crystallization.
  This phenomenon involves in turn a highly non-trivial interdependence between the molecular level of bond-orientation-dependent nucleation, and the macroscopic level, where the temperature-dependent rheology generates stretch of entire chain segments \cite{Graham09, Troisi17, Nicholson19, Moghadam19, Read20}. 
  Remarkably, nature has found a way to control robustly  the flow-induced self-assembly of silk from an intrinsically disordered state (a solution of random-walk polymers) prior to forming high-performance fibers under flow at ambient conditions   \cite{Asakura83, Asakura84, ZhaoC01, Holland12a, Asakura15, Laity15, Laity16, Schaefer20}).
  Key to achieving the final properties is that silk is processed in semi-dilute aqueous conditions \cite{Holland12a}, where nucleation can be induced through the stretch-induced disruption of the solvation layer \cite{Dunderdale20}. 
  How sufficient polymer stretch can be achieved in a limited time under modest flow conditions has so far remained unexplained.
  Recent work has shown that microscopic chain stretch and the consequent macroscopic strain hardening is triggered by a small number of  calcium bridges \cite{Koeppel18, Schaefer20} that act as `sticky' reversible intermolecular crosslinks akin to those in synthetic `sticky polymers' \cite{KramerBook88, Leibler91, Weiss91, Annable93, Colby98, Weiss07, Seiffert12, Hackelbusch13, ZhangZ18}.
For this class of molecules, a molecular understanding of the non-linear rheology and crystallization of `sticky polymers' has so far relied on computationally expensive (coarse-grained) molecular dynamics simulations \cite{ ChenQ16, Tomkovic18, Tomkovic19, Zuliki20, Cui18, Read20}. Simpler molecular models coarse-grained at the level of entanglements, but able to capture the vital slow processes, remain absent.

In the present work, we address this need by following the central idea by  de Gennes of replacing the many-chain problem with a single chain in a tube-like confinement imposed by its environment of entanglements \cite{deGennes71}, and solve the Brownian dynamics of the chain in $1$D \cite{DoiEdwards, Likhtman02}.
This approach is simple yet powerful, and has led to the development of widely applied finite-element solvers (\cite{Likhtman03, Boudara20, Collis05}), a physical explanation for the  (apparent) $3.4$ power dependence of the relaxation time of polymer melts on the molecular-weight \cite{Doi83}, and a comprehensive understanding of the rich non-linear rheology of (bimodal) polymer blends \cite{Graham03, Auhl09}.
The ingredient that we add in this letter is a `sticky-reptation model' description \cite{Green46} for the temporary binding of associating monomers to the tubular environment developed for full non-linear flows. The model shares some structural similarities with early `transient network' approaches to polymer melt and solution rheology, also demonstrating a hitherto unrecognised feature of those models

%\section{Model}

The starting point of our contribution is to consider a chain consisting of $N$ Kuhn segments with length $b$, and $Z_\mathrm{e}$ entanglements (hence, with tube diameter $a=b(N/Z_\mathrm{e})^{1/2}$).
The configuration of the chain is given by the spatial coordinates $R_i$ of monomers $i=1,\dots,N$ along the curvilinear direction along the tube, which evolve with time according to the Langevin equation \cite{Doi83, Likhtman02, Graham03}
\begin{equation}
\zeta \frac{\partial R_i}{\partial t}
=
 \left(
  \frac{3\kT}{b^2}\frac{\partial^2 R_i}{\partial i^2}
+ f_i \right)(1-p_i) + \dot{\varepsilon} \zeta (R_i - {R}_{\mathrm{C.M.}}),
\label{eq:Langevin1D}
\end{equation}
with $\partial R/\partial i = a$ at $i=1$ and at $i=N$, $\zeta$ the monomeric friction, $\kT$ the thermal energy, and $f_i$ a stochastic force given by the equipartition theorem  
\begin{equation}
\langle f_i(t)\rangle=0; \,\,\langle f_i(t)f_{i'}(t')\rangle = 2\kT \zeta \delta(i'-i)\delta(t'-t).
\end{equation}
In the absence of stickers, this equation predicts the Rouse diffusivity \cite{DoiEdwards}
\begin{equation} 
  D_\mathrm{R}=\frac{a^2}{3\pi^2\tau_\mathrm{e}Z_\mathrm{e}}=\frac{\kT}{\zeta N} \label{eq:RouseDiffusivity}
\end{equation}
 and the variance of quiescent contour-length fluctuations $\langle |R_N-R_1|^2\rangle = a{Z_\mathrm{e}}/3$.
${R}_{\mathrm{C.M.}}$ is the center of mass of the chain, and the strain rate, $\dot{\varepsilon}$, is in one spatial dimension equivalent to the strain rate in the GLaMM model \cite{Graham03}.

To model the binding and unbinding of monomers to the environment, we introduce a stochastic state variable $p_i(t)$, which takes values of either zero or unity for each monomer $i$.
If it is zero, the monomer is free to diffuse and respond to the drag exerted by the flow field, as well as to respond to the stress within the polymer, including at least some relaxation of stress in segments it adjoins.
However, if the value of the state variable is unity, the monomer is kinetically trapped by its environment and is unable to diffuse or to respond to the differences in chain tension exerted by the polymer.
Hence, the closed sticker advects with the background flow.
While the state variable of regular (non-sticky) monomers is always zero, the state variable for sticky monomers can either be zero (`the sticker is open'), or unity (`the sticker is closed').  
In our simulations, the opening and closing of the stickers is simulated using a simple stochastic algorithm, where at every time step $\Delta t$ a sticker is opened with probability $k_{i,\mathrm{open}} \Delta t$ or closed with probability $k_{i,\mathrm{close}} \Delta t$.

By defining the rules by which $p_i$ may switch between the `open' and `closed' states, copolymers with arbitrary monomer sequences may be modeled. Here, we consider chains with $N-Z_\mathrm{s}$ monomers that are non-sticky (for these monomers $k_{i,\mathrm{close}}=0$ and $p_i=0$ at all times) and $Z_\mathrm{s}$ sticky monomers that may switch state using rates $k_{\mathrm{close}}$ and $k_{\mathrm{open}}$. 
%If the opening and closing of the sticker is fast, the stochastic factor $p_i(t)$ may be replaced by $ p $, which is the time- or ensemble-averaged value of $p_i$. This renormalises the monomeric friction for the sticker, $\zeta_\mathrm{s} =  \zeta/(1- p)$.
%At increasing strain rate, however, the association rate becomes  
%In this regime, the opening and closing of the stickers has to be modeled explicitly.
The opening rate is related to the rheological sticker lifetime, $\tau_\mathrm{s}=k_\mathrm{open}^{-1}$ \cite{Leibler91, ChenQ16, ZhangZ18, Tomkovic18, Tomkovic19, Zuliki20, Schaefer20}, and the closing rate is given by $k_\mathrm{close}=k_\mathrm{open}p/(1-p)$, with $p$ the time- or ensemble-averaged fraction of closed stickers.
Here, we ignore the underlying dissociation-association or bondswap mechanisms that determine the concentration-dependence of the opening and closing rates \cite{Smallenburg13, Ciarella18}, and view $p$ and $\tau_\mathrm{s}$ as free model parameters.

%In this mechanism, an open sticker may close in presence of a closed sticker pair through the rate $k_\mathrm{bondswap}[S]p$ with $[S]$ the overall sticker concentration (note that $p$ impliticly depends on $[S]$, but is independent of bond swapping \footnote{The fraction of closed stickers $p$ is given by $K[S]=p/(1-p)^2$ with $[S]$ the overall sticker concentration and $K=k_\mathrm{ass}/k_\mathrm{diss}$ an equilibrium constant with $k_\mathrm{ass}$ and $k_\mathrm{diss}$ the association and dissociation rates, respectively.}), and a close sticker may open in the presence of an open sticker through the rate $k_\mathrm{bondswap}[S](1-p)$.

\begin{figure}[ht!]
  \centering
  \includegraphics*[width=8cm]{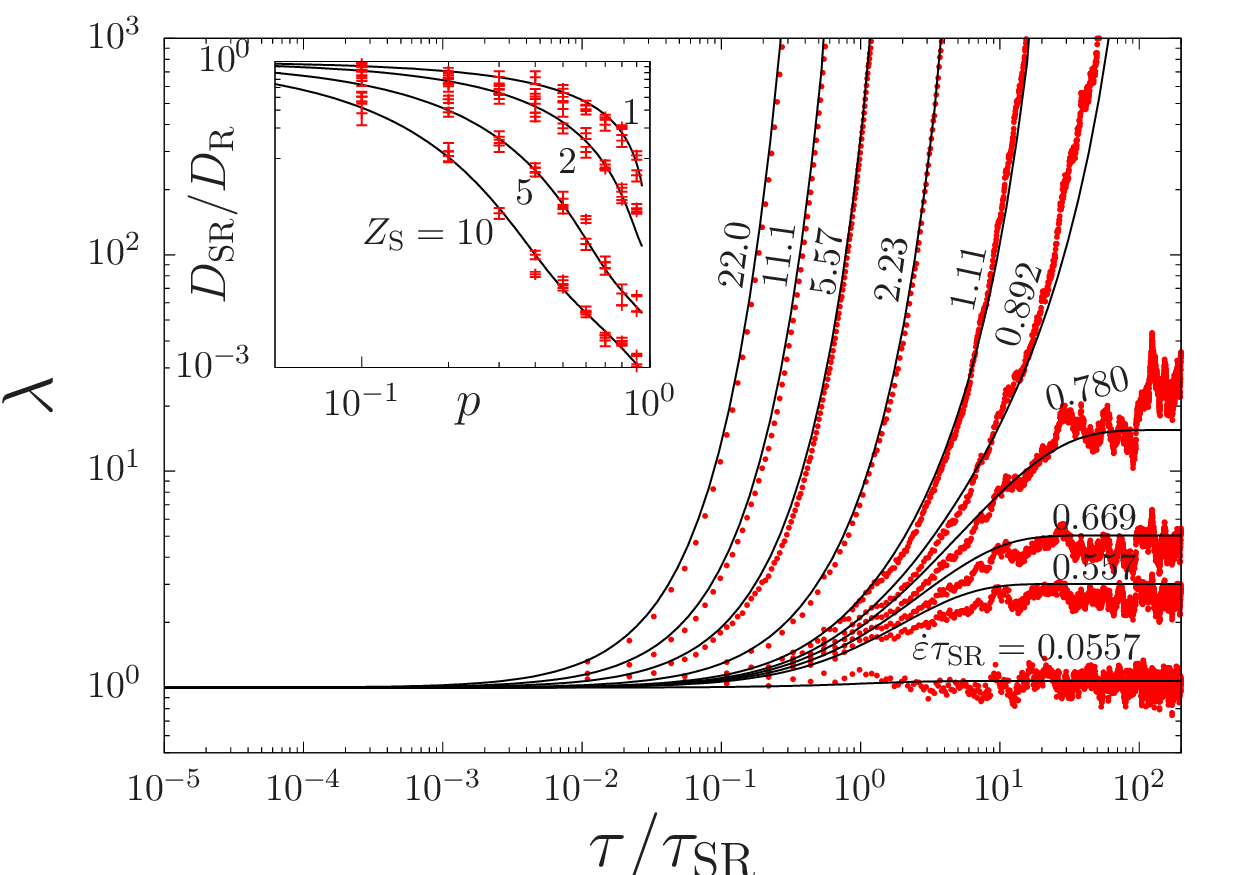}
  \caption{
Comparison between the stretch ratio $\lambda$ of a sticky polymer ($Z_\mathrm{e}=Z_\mathrm{s}=10$, $\tau_\mathrm{s}=10^4\tau_\mathrm{e}$, $p=0.95$, $Z_\mathrm{s}=10$) against time $t$ in units of the sticky Rouse time $\tau_\mathrm{SR}$. The sticky Rouse time is $\tau_\mathrm{SR}=[D_\mathrm{R}/D_\mathrm{SR}]\tau_\mathrm{R}$ with $D_\mathrm{R}$ the bare Rouse diffusivity, $\tau_\mathrm{R}=\tau_\mathrm{e}Z_\mathrm{e}^2$ the bare Rouse time and $D_\mathrm{SR}$ the sticky diffusivity (see inset).
In the main panel, the symbols are obtained by averaging over five Brownian dynamics simulations with different random number seeds, and the lines represent the single-mode model in Eq.~(\ref{eq:SingleModeModel}). 
The flow rate is increased from $\dot{\varepsilon} = 0.056 \tau_\mathrm{SR}^{-1}$ to $22.3  \tau_\mathrm{SR}^{-1}$ in logarithmic steps.
%The flow rate is increased in a $1-2-5$ series from $\dot{\varepsilon} = 0.00005$ to $0.02$ in units of the bare Rouse time $\tau_\mathrm{e}Z_\mathrm{e}^2$.  DSR=3.032970e-06; DR=1.0/(3*PI*PI*ZE)=0.0033808 --> tauSR=1114.7*tauR
The inset shows consistence of the simulated sticky-Rouse diffusivity (symbols; averaged over $25$ random number seeds) with the sticky-reptation model of Leibler \emph{et al} \cite{Leibler91}.
 } \label{fig:StretchingZS10}
\end{figure}

%The overall effective dissociation time, $\tau_\mathrm{s}=k_\mathrm{open}^{-1}$, measured in the linear rheology \cite{Leibler91, ChenQ16, ZhangZ18, Tomkovic18, Tomkovic19, Zuliki20, Schaefer20} is
%set by the opening rate 
%\begin{equation}
%  k_\mathrm{open} = k_\mathrm{diss} + k_\mathrm{bondswap}[S](1-p).
%\end{equation} 
%In the present work, we emphasise the importance of the sticker closing rate  
%\begin{equation}
%  k_\mathrm{close} = k_\mathrm{diss}\frac{p}{1-p} + k_\mathrm{bondswap}[S]p.
%\end{equation}
%(note that $k_\mathrm{close}(1-p)=k_\mathrm{open}p$, as it should) 
%These equations reveal that in asymptotically low or high fractions of bound stickers the dynamics are dominated by the association-dissociation mechanism, while at intermediate fractions bondswapping dominates the dynamics if $k_\mathrm{bondswap}[S]/k_\mathrm{diss} > 1$.

%With this mechanism in mind, in the following we will consider the fraction of bound stickers and the opening and closing rates as  independent properties of the stickers, and explore the emerging dynamic regimes of polymer diffusion and stretching. 
%We explore these regimes by solving the Langevin equation for various chain lengths in units of $Z_\mathrm{e}$, various numbers of stickers $Z_\mathrm{s}$ distributed on the backbone of the chain. 
%(In order to do so, we follow the approach in \cite{Likhtman02} and simulate the Langevin equation at various levels of coarse graining, i.e., by describing a chain of fixed $Z_\mathrm{e}$ entanglements using a varying number of beads.)

We have benchmarked our model in the absence of flow using the Likhtman-McLeish model for linear non-sticky polymers (results not shown here) and using the sticky-Rouse diffusivity, $D_\mathrm{SR}=D_\mathrm{SR}(Z_\mathrm{e}, \tau_\mathrm{e}, Z_\mathrm{s}, \tau_\mathrm{s}, p)$ as calculated by Leibler et al. \cite{Leibler91} (see the inset of Figure~\ref{fig:StretchingZS10}).
For the non-linear dynamics of sticky polymers, so far no comparisons between analytical predictions with simulations or experiments have been reported.
The first strategy to address this is to evaluate how well a DEMG-type single-mode approximation performs, with chain friction renormalized by averaging over the stochastic sticker dynamics: 
\begin{equation}
  \frac{d\lambda}{dt}=\dot{\varepsilon}\lambda + \frac{1}{\tau_\mathrm{SR}}(1-\lambda) \label{eq:SingleModeModel}
\end{equation}
where the stretch ratio, $\lambda \equiv (R_N-R_1)/Z_\mathrm{e}$, is presumed to be uniform over the backbone of the chain. The extension rate is proportional to the stretch ratio itself. The retraction rate is determined by $1-\lambda$ (in the absence of flow, $\lambda=1$ at steady state) and by the sticky-Rouse time, $\tau_\mathrm{SR}\equiv [D_\mathrm{R}/D_\mathrm{SR}]\tau_\mathrm{S}$.    
In the main graph of Figure~\ref{fig:StretchingZS10}, we present comparison between this simple approximation and our simulations,  (the approximations inherent in the DEMG require that the simulation time be divided by a factor $1.2$ to result in the close agressment shown).
This confirms that the intuitive `sticky Weissenberg number' for the stretch transition is $\mathrm{Wi}=\dot{\varepsilon}\tau_\mathrm{SR}$.
For $\mathrm{Wi}>1$ an exponential runaway stretch emerges as expected. 
In contrast to non-sticky polymers, however, we will argue that the stress and fluctuation in stretch may diverge \emph{below} this stretch transition when the pre-averaging approximation inherent in DEMG is avoided. 

While non-sticky polymers in steady state show a Gaussian stretch distribution with a width that is determined by the (effective) number of entanglements, we have observed rather large stretch fluctuations for the sticky polymer at extension rates of the order of, but below, the critical value. 
Indeed, the symbols in Figure~\ref{fig:StretchingZS10} are averaged over five simulations for a chain with $10$ stickers which are on average closed a fraction $p=0.95$ of time.
For simulations with $p<0.9$ these fluctuations become much larger and difficult to distinguish graphically.
Indeed, while the mean stretch is finite, the fluctuations in stretch diverge above a certain flow rate \emph{below} the stretch transition.    

\begin{figure}[ht!]
  \centering
  \includegraphics*[width=8cm]{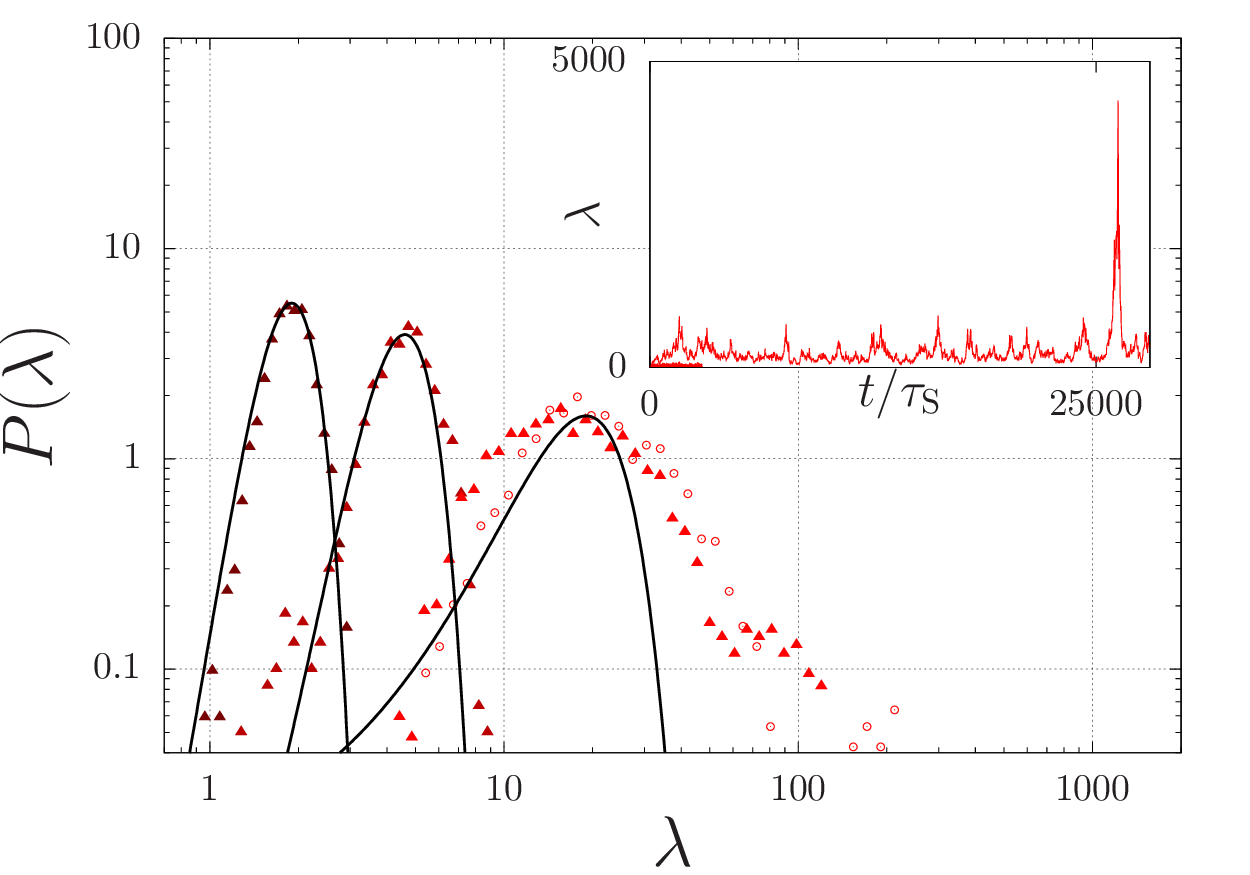}
  \caption{
The steady-state probability distribution, $P(\lambda)$, is plotted against the stretch ratio, $\lambda$. The symbols are obtained from the steady-state simulations of Fig.~\ref{fig:StretchingZS10} at the flow rates ($\dot{\varepsilon}\tau_\mathrm{SR}=0.446,\,0.668$ and $0.780$; the curves are Gaussian fits. 
For an increasing flow rate, the high-stretch tail is no longer Gaussian but becomes a power law, $P(\lambda)\propto \lambda^{-\nu}$. The inset shows the stretch ratio against time, and visualizes how this distribution includes `rare events' of enormous chain stretch. For a sufficiently large flow rate, $\nu$ decreases. If $\nu >2$, the mean value of $\lambda$ is finite (as it should in steady state); however, if also $\nu \leq 3$, the fluctuations in stretch, characterized by
the expectation value of $\lambda^2$, diverge.
 } \label{fig:StretchDisitributionZS10}
\end{figure}

For three flow-rates of Figure~\ref{fig:StretchingZS10} we have plotted the stretch distribution, $P(\lambda)$, in Figure~\ref{fig:StretchDisitributionZS10}.
For small flow rates, the stretch distribution is Gaussian, $\ln P(\lambda) \propto (1-\lambda)^2$ (solid curves), as in the quiescent state.
However, for increased flow rates deviations emerge in the high-$\lambda$ tail of the distribution.
Importantly, the polymer stretch may resemble the mean stretch for long times compared to the sticky-Rouse time, and only in  `rare events' the stickers may remain closed sufficiently long for the stretch to reach deep into the tail of the distribution (see inset).  
%We argue these events may be related to slow processes in extensional flow, such as the crystallisation of spider silk in flows as slow as a few micron per second \cite{Holland12B}.
%We speculate that gelation and crystallisation of silk under these conditions may be responsible for the long relaxation times observerd in stop-flow experiments \cite{Laity16B}.

In the following, we will simplify the problem using a `sticky dumbbell model' to explore and clarify the underlying causes of the power-law tail in the stretch distribution, and explore how it can be tuned by the flow rate.
This minimal model that captures the essential physics is equivalent to a single polymer strand either attached to the bulk deformation at both ends (the `closed’ state) or free to relax (the `open’ state).
The rate by which the polymer switches between the two states is given by the usual opening and closing rates. 
We can now address the development of stretch under extensional flow through a pair of coupled  partial differential equations for the time-dependent stretch distributions $P_\mathrm{o}(t,\lambda)$ and $P_\mathrm{c}(t,\lambda)$ for each state using the master equation
\begin{align}
  \frac{\partial P_\mathrm{c}}{\partial t} &= -\frac{\partial}{\partial \lambda}\left[P_\mathrm{c}\dot{\varepsilon}\lambda\right]-k_\mathrm{open}P_\mathrm{c}
 -k_\mathrm{close}P_\mathrm{o}, \nonumber\\
  \frac{\partial P_\mathrm{o}}{\partial t} &= -\frac{\partial}{\partial \lambda}\left[P_\mathrm{o}\left(\dot{\varepsilon}\lambda+\frac{1-\lambda}{\tau_\mathrm{R}}\right)\right]+k_\mathrm{open}P_\mathrm{c}
-k_\mathrm{close}P_\mathrm{o}. \label{eq:masterequation}
\end{align}
Note that this evolution equation invokes a single-mode approximation and ignores thermal fluctuations: the stretch distribution emerges from the coupling between a `closed' state in which the polymer is stretched and the `open' state in which it can retract. Under strong flow conditions, the effective driving noise is completely dominated by the stochastic state-switching, with thermal noise negligible.

\begin{figure}[ht!]
  \centering
  \includegraphics*[width=8cm]{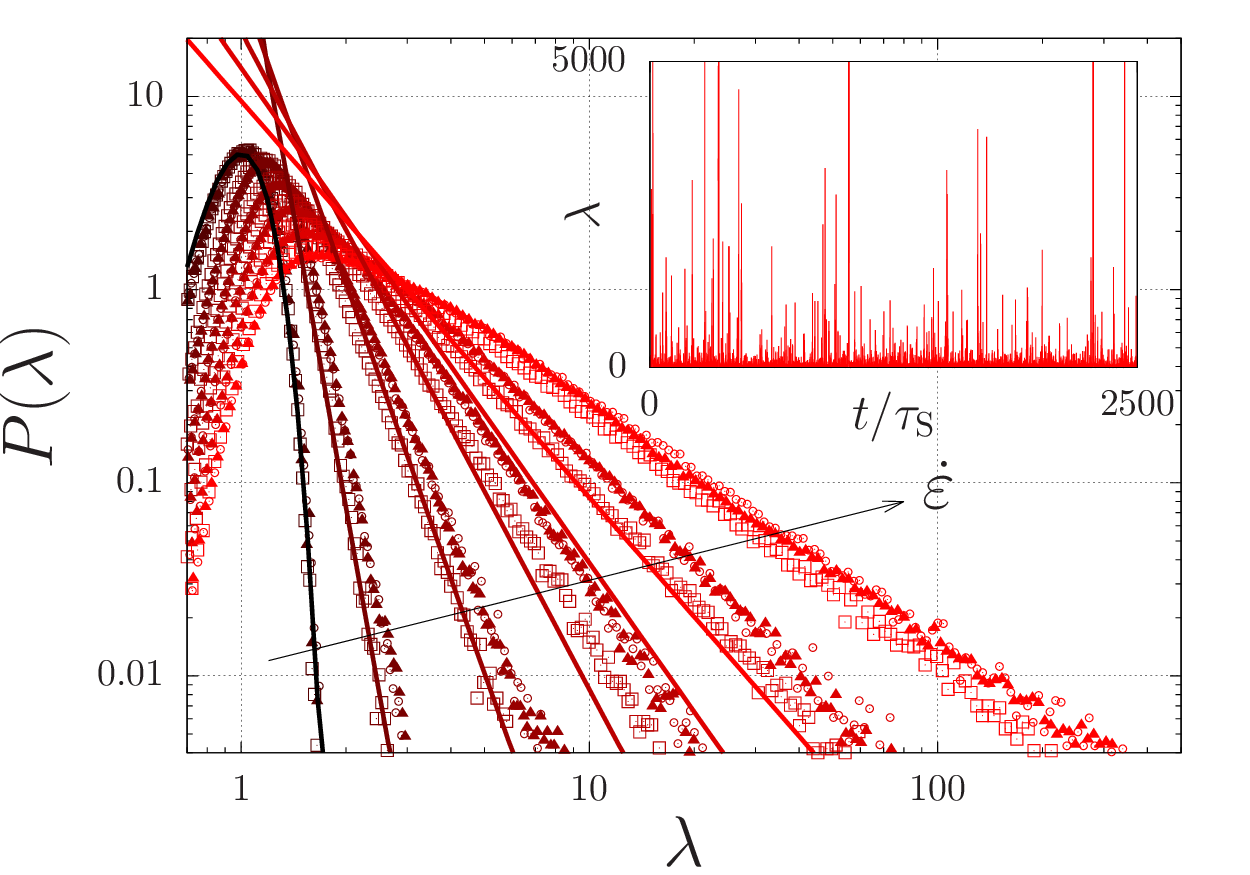}
  \caption{
The power-law stretch distribution, $P(\lambda)\propto \lambda^{-\nu}$ for large $\lambda$, observed in Fig.~\ref{fig:StretchDisitributionZS10} is replicated analytically in a sticky dumbbell model for a sticky polymer ($Z_\mathrm{e}=10$, $p=0.9$, $\tau_\mathrm{s}=1000\tau_\mathrm{e}$), which has two stickers near the end of the chain that are simultaneously either open or closed (lines). The black curve is the Gaussian stretch disitribution under quiescent conditions. In linear steps, the flow rate is increased up to $\dot{\varepsilon}\tau_\mathrm{R}=0.05$. The squares, circles and triangles were obtained in simulations with $6$, $12$ and $36$ beads, respectively. 
For small flow rates, where $\nu<3$, the simulated power-law tails of $P(\lambda)$ (symbols) are in agreement with Eq.~(\ref{eq:PowerLaw}). For higher flow rates the simulated stretch of the multi-bead chains exceeds the single-mode theory. The inset shows the transient behavior of the simulation with $\dot{\varepsilon}\tau_\mathrm{R}=0.05$. 
 } \label{fig:StretchDumbbell}
\end{figure}

We calculate the steady-state stretch distribution at strong stretch by setting the left-hand side of Eq.~(\ref{eq:masterequation}) to zero and taking $\lambda\gg 1$. The result can be solved analytically since in these conditions the differential system becomes homogeneous. We therefore find the power-law relation
\begin{equation}
  P(\lambda) \propto \lambda^{-\nu},
\end{equation}
with the exponent given in terms of the three dimensionless parameters of the system, $p$, $\dot{\varepsilon}\tau_\mathrm{R}$, $\tau_\mathrm{R}/\tau_\mathrm{s}$ by
\begin{equation}
  \nu=1+\frac{1}{1-\dot{\varepsilon}\tau_\mathrm{R}}\frac{p}{1-p}\frac{\tau_\mathrm{R}}{\tau_\mathrm{s}}-\frac{1}{\dot{\varepsilon}\tau_\mathrm{s}}.
\label{eq:PowerLaw}
\end{equation}
We compare this power-law to our sticky dumbbell simulations in Figure~\ref{fig:StretchDumbbell}.

\begin{figure}[ht!]
  \centering
  \includegraphics*[width=8cm]{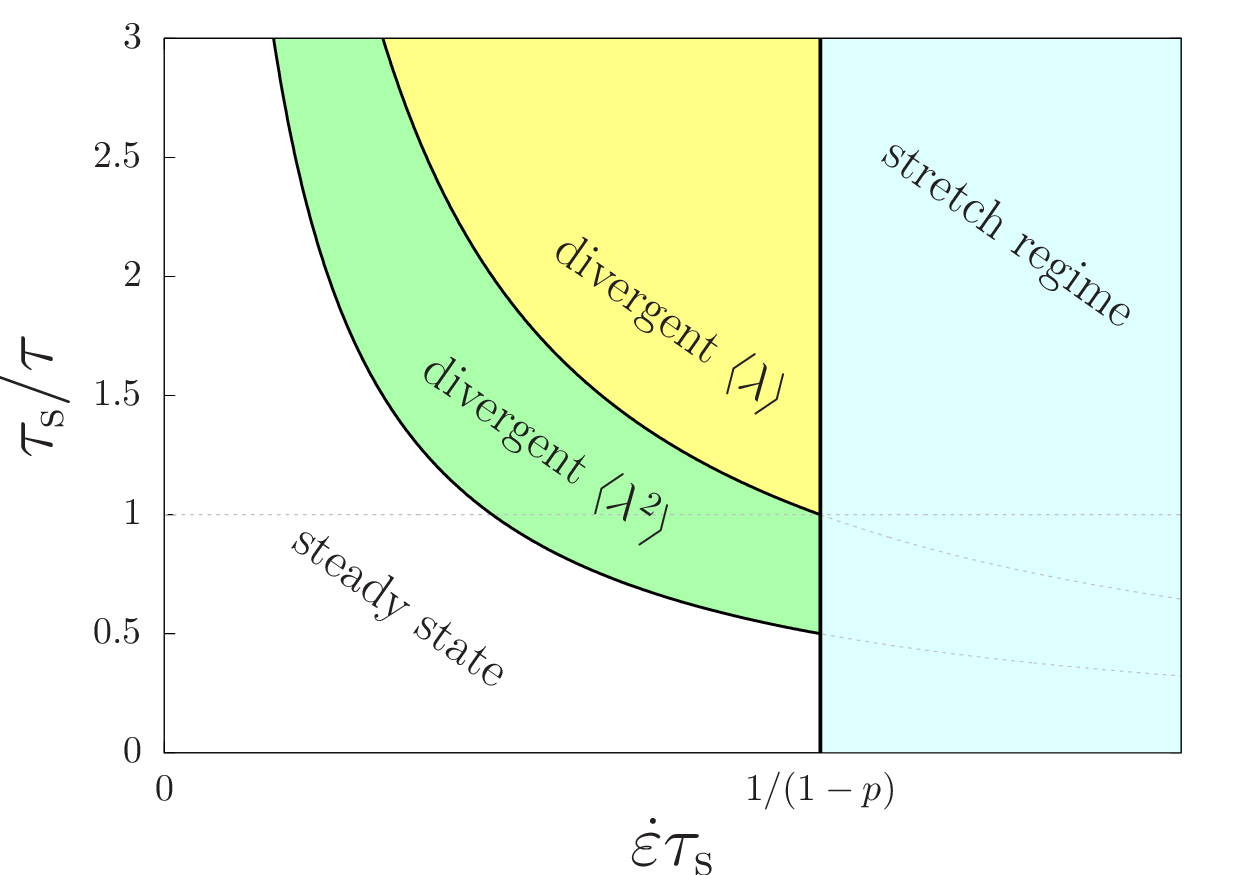}
  \caption{
State diagram of a chain with a sticker at each chain. For a short sticker lifetime, $\tau_\mathrm{s}<\tau_\mathrm{R}/2$, polymer stretching takes place above the transition at the Weissenberg number $\dot{\varepsilon}\tau_\mathrm{R} = 1/(1-p)$, with $p$ the time-averaged fraction of closed stickers and $\tau_\mathrm{R}$ the bare Rouse time.
For a longer sticker lifetime, new regimes emerge, in the first of which the chains have a finite mean stretch, $\langle \lambda\rangle$, but with a divergence in fluctuations, i.e., $\langle \lambda^2\rangle \rightarrow \infty$. In this regime, power-law stretch distributions emerge (see Figure \ref{fig:StretchDumbbell}).
 } \label{fig:PhaseDiagram}
\end{figure}

For sufficiently small flow rates, we find a reasonable agreement between our multibead simulations and the analytical approximation for the simple sticky dumbbell (under these conditions, $\nu>3$).
When $\nu$ approaches a value $3$ (this occurs at $(1-p)\dot{\varepsilon \tau_\mathrm{R}}\approx \tau_\mathrm{R}/(2\tau_\mathrm{s})$) the discrepancies become larger. This is not a coincidence: if $\nu=3$ the magnitude of the fluctuations diverge, $\langle \lambda^2 \rangle\rightarrow \infty$. 
Indeed, near this condition the variations in stretch are enormous and the single-mode approximation breaks down for multibead chains.
We confirm this using simulations with a variable number of beads per chain: for fewer beads the higher-order Rouse modes are removed and the simulations show a better agreement with  the single-mode approximation.
Although the fluctuations diverge for $\nu=3$, the mean  $\langle \lambda \rangle $ remains finite as long as $\nu\leq 2$ (the equality holds approximately when $(1-p)\dot{\varepsilon \tau_\mathrm{R}}\approx \tau_\mathrm{R}/\tau_\mathrm{s}$). For even larger flow rates, i.e., for $\nu\leq 1$ (at $(1-p)\dot{\varepsilon \tau_\mathrm{s}}=1$) the stretch distribution can no longer be normalized and true runaway stretch emerges.
These various regimes are displayed in Figure~\ref{fig:PhaseDiagram} in terms of the dimensionless parameters of the system.
Note that the stress is $\sigma\propto (1-\lambda)^2$ and the tail of the stress distribution is  $P(\sigma)\propto \lambda^{-\nu/2}$: the mean stress diverges for $\nu\leq 4$ and its variance diverges for $\nu\leq 6$.
 
The single-mode toy model clarifies the route through which the divergent fluctuations arise. Crucially, when a stretched strand is freed from the network, it may not relax entirely before reattachment (this effect is ignored in classical treatments of transient network models, which in consequence overlook the strong stochastic fluctuations they physically imply). Such continuous  interchange between convecting and relaxing strands, together with the occurrence of longer-than-average attachment times for some segments, allow the exploration of very large chain stretches in steady-state.

To illustrate the potential consequences of this effect, we consider nucleation rates in steady-state extensional flow, assuming that polymer crystal phase may nucleate around chains beyond a critical stretch ratio $\lambda_\ast$ \cite{Graham09}. 
Assuming that the chain is relaxed prior to sticker closing at time $t=0$, its stretch ratio develops as $\lambda(t)=\exp(\dot{\varepsilon}t)$ until it closes at a time $\tau_\mathrm{open}$.
This time is drawn from the probability distribution $p(\tau_\mathrm{open})=\tau_\mathrm{s}^{-1}\exp(-\tau_\mathrm{open}/\tau_\mathrm{s})$, so the probability that the critical stretch is reached is $p_\ast = \lambda_\ast^{-1/\dot{\varepsilon \tau_\mathrm{s}}}$.
The probability that  $\lambda_\ast$  is not reached after $n$ attempts is $(1-p_\ast)^n$, and therefore the expected number of attempts needed is  
\begin{equation}
  \langle n\rangle = \frac{\sum_{n=1}^\infty n(1-p_\ast)^n}{\sum_{n=1}^\infty (1-p_\ast)^n}
  = \lambda_\ast^{1/(\dot{\varepsilon}\tau_\mathrm{s})}.
\end{equation}
An attempt occurs, on average, after time intervals $1/k_\mathrm{open}+1/k_\mathrm{close} = \tau_\mathrm{s}/p$.
If the number density of chains is $\rho$, then combining these results gives 
\begin{equation}
  J  = \frac{\rho p}{\tau_\mathrm{s}} \lambda_\ast^{-1/(\dot{\varepsilon}\tau_\mathrm{s})}.
\end{equation}
as the extension-rate-dependent nucleation rate per volume.

In conclusion, we have numerically solved the stochastic Langevin equation of an aligned entangled sticky polymer in an effective medium and in extensional flow.
We have found that the  stretch transition is determined by the sticky-Rouse time and that the early stretch transients are well described using a single-mode approximation.
Below the transition, we have identified a steady-state regime where the time- or ensemble-averaged stretch distribution has a power-law tail that renders large stretches much more likely than in the usual contour-length-fluctuation-dominated Gaussian distribution. 
We have shown that this behavior originates from a stochastic coupling between stretching and relaxation states of the polymer stretch distribution.
We expect that these insights provide new means to calculate nucleation rates for polymer crystallization in extensional flow.
It also provides an example of one of a family of driven, stochastic, systems in which a divergent and scaling structure of fluctuations arises, not just at a single critical point, but within a large region of state space, and with a universal critical exponent replaced by a family, dependent on the degree of forcing.
 
% For the dumbbell we will calculate a nucleation rate.
%At the end of this letter, we will argue that the physics is robust and carries over to the multi-sticker chain. 
%We will provide the guidelines to generalize the model; the detailed analysis will be presented in forthcoming work.

\begin{acknowledgments} % RevTeX
%\begin{acknowledgement}     % achemso 
This research was funded by the Engineering and Physical Sciences Research Council [grant number EPSRC (EP/N031431/1)]. Jorge Ram\'{i}rez is thanked for  sharing Alexei Likhtman's Brownian dynamics code and Richard Graham is thanked for sharing his GLaMM code; both codes helped us to benchmark our simulation software. Chris Holland and Pete Laity are thanked for useful and encouraging discussions.
%\end{acknowledgement}       % achemso
\end{acknowledgments}   % RevTeX

\bibliography{references} % Linux
%\printbibliography
%\end{widetext}
%\lipsum[1-3]
\end{document}